\documentclass[twocolumn,showpacs,preprintnumbers,amsmath,epsfig]{revtex4}

\usepackage{graphicx}
\usepackage{graphics}
\usepackage{dcolumn}
\usepackage{bm}

\begin{document}


\title{Spin dynamics of heterometallic Cr$_{7}$M wheels (M = Mn, Zn,
Ni) probed by
inelastic neutron scattering}

\author{R. Caciuffo and T. Guidi}

\affiliation{Dipartimento di Fisica ed Ingegneria dei Materiali e del
Territorio, Universit\`a Politecnica delle Marche, Via Brecce Bianche, I-60131
Ancona, Italy}

\author{G. Amoretti, S. Carretta, E. Liviotti, and P. Santini}

\affiliation{Dipartimento di Fisica, Universit\`a di Parma, I-43100
Parma, Italy}

\author{C. Mondelli}

\affiliation{Istituto Nazionale per la Fisica della Materia, and
Institut Laue Langevin, Bo\^{i}te Postale 220 X, F-38042 Grenoble
Cedex, France}

\author{G. Timco, C. A. Muryn  and R.E.P. Winpenny}

\affiliation{Department of Chemistry, The University of Manchester,
Oxford Road, Manchester, M13 9PL, United Kingdom}

\begin{abstract}
Inelastic neutron scattering has been applied to the study of the
spin dynamics of Cr-based antiferromagnetic octanuclear rings
where a finite total spin of the ground state is obtained by
substituting one Cr$^{3+}$ ion (s = 3/2) with Zn (s = 0), Mn (s =
5/2) or Ni (s = 1) di-cations. Energy and intensity measurements
for several intra-multiplet and inter-multiplet magnetic
excitations allow us to determine the spin wavefunctions of the
investigated clusters. Effects due to the mixing of different spin
multiplets have been considered. Such effects proved to be
important to correctly reproduce the energy and intensity of
magnetic excitations in the neutron spectra. On the contrary to
what is observed for the parent homonuclear Cr$_{8}$ ring, the
symmetry of the first excited spin states is such that
anticrossing conditions with the ground state can be realized in
the presence of an external magnetic field. Heterometallic
Cr$_{7}$M wheels are therefore good candidates for macroscopic
observations of quantum effects.
\end{abstract}


\maketitle

\section{Introduction}

Magnetic wheels are polynuclear molecular clusters with a
ring-shaped cyclic structure and a dominant antiferromagnetic (AF)
coupling between nearest-neighbor ions. For an even number of spin
centers, the ground state is a singlet and the excitation spectrum
is characterized by rotational and spin-wave like bands
\cite{Schnack00,Waldmann01a,Waldmann03a}. Heterometallic rings
with S$\neq 0$ can be obtained from an S = $0$ homonuclear ring by
chemical substitution of one or two magnetic centers
\cite{Larsen03a,Larsen03b}. Theoretical calculations suggest that
such systems could exhibit interesting quantum-coherence phenomena
\cite{Meier03a,Meier03b} and are therefore of considerable
interest. Indeed,  the replacement of a magnetic ion in a cyclic
structure allows one to modify the topology of the exchange
interactions, that in turn plays a key role in determining the
macroscopic behavior of the system
\cite{Carretta03b,Waldmann04a,Carretta04a}.

Here we report the results of inelastic neutron scattering (INS)
experiments on heterometallic AF rings, and we show that the spin
level sequence and dynamics are substantially modified with
respect to those of the parent compound. The investigated wheels
are derived from the spin-compensated neutral
[Cr$_{8}$F$_{8}$(O$_{2}$CCMe$_{3}$)$_{16}$] ring
\cite{vanslageren02} by substitution of one divalent cation (M =
Mn, Zn, Ni) for a trivalent Cr ion \cite{Larsen03a}. Due to the
difference between the Cr$^{3+}$ spin (s = $3/2$) and the spin of
the dication, the ground state of the so-obtained Cr$_{7}$M wheels
has a non-zero total spin (S = $1/2$, $1$ and $3/2$ for M = Ni, Mn
and Zn respectively). Recent investigations have shown that,
within this Cr$_{7}$M family, S = $3/2$ rings exhibit
magnetocaloric effects that could be exploited below T $\sim$ $2$
K \cite{Affronte04}, whilst Cr$_{7}$Ni could be a suitable
candidate for the physical implementation of qubits
\cite{Troiani04}. The suggested opportunities are linked to
properties that are very sensitive to the spin energy spectrum and
to the composition of the spin wavefunctions. A reliable
determination of these quantities is therefore relevant, and INS
is known to be the most appropriate technique to address this
problem \cite{Caciuffo98,Carretta03a,Guidi04a,Carretta04b}.

Our INS experiments on Cr$_{7}$M provide energy and intensity for
several transitions between the anisotropy-split lowest spin
multiplets. The analysis of the data leads to an accurate
determination of the main features of the microscopic
intra-cluster interactions, including nearest-neighbor isotropic
exchange, zero-field splitting and dipole-dipole interaction
parameters (inter-cluster interactions are very weak, of the order
of 10 neV). The results obtained indicate that anticrossing
conditions between the ground state and excited low-lying energy
levels with the same symmetry can be met in Cr$_{7}$M, by applying
a suitable magnetic field. This is at variance with the situation
in the parent Cr$_{8}$ ring, where ground state crossings involve
states with different symmetry \cite{Carretta03a}, as confirmed by
the vanishingly small level repulsion observed by heat capacity
measurements \cite{Affronte03}. The occurrence of anticrossing
conditions suggests that Cr$_{7}$M wheels are good candidates for
the macroscopic observation of quantum phenomena arising from
S-mixing, such as the quantum fluctuation of the magnitude of the
total spin \cite{Carretta03b,Waldmann04a}.

A correct interpretation of the experimental spectra requires a
generalization of the orientation-averaged INS cross-sections for
polycrystals
\cite{Borras99}, as proposed
by Waldmann \cite{Waldmann03b}.

\section{Experimental details}

\begin{figure}
\includegraphics[width=7cm]{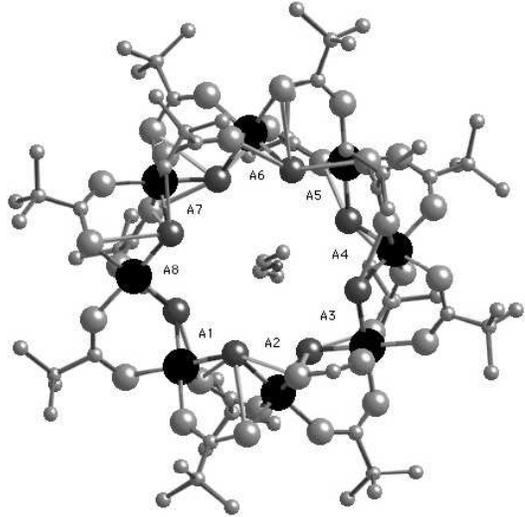}
\caption{The molecular structure of the Cr$_{7}$M (M = Mn, Ni, Zn)
magnetic ring, with hydrogen atoms omitted for clarity. The A ions
(A = $7/8$ Cr and $1/8$ M at each site) are represented by large
dark circles; fluorine positions are indicated by small dark
circles, the oxygen and carbon atoms are large and small gray
circles, respectively. The cation is found H-bonded inside the
cavity.} \label{Cr7Mstructure}
\end{figure}

Deuterated  micro-crystalline samples of (C$_2$D$_5$)$_2$NH$_2$
[Cr$_7$MF$_8$(O$_2$(C$_5$D$_9$))$_{16}$ ] (M$^{2+}$ = Ni, Mn, Zn)
have been prepared according to a slightly modified literature
procedure \cite{Larsen03a,Larsen03b}, by dissolving chromium
fluoride tetrahydrate (Aldrich) in a mixture of trimethyl-{\it
d$_9$}-acetic acid and diethyl-{\it d$_{10}$}-amine  (98$\%$
deuterated, Aldrich) before adding an excess of the second metal
salt  (nickel carbonate hydroxide tetrahydrate, manganese chloride
tetrahydrate or basic zinc carbonate). The preparation of
trimethyl-{\it d$_9$}-acetic acid starting from acetone-{\it
d$_6$} was adapted from standard methods \cite{Furniss89}. All
compounds were purified additionally on a silica gel column using
toluene as the eluent \cite{Larsen03a} and finally crystallized
from a mixture of pentane/acetone by evaporation of the mixture of
solvents at 313-318 K.  X-ray diffraction analyses for all three
compounds showed that crystals don't have any solvent molecules in
the crystal lattice. The observed Bragg peaks could be indexed in
the tetragonal {\it P}4 space group, with lattice parameters at
100 K of  {\it a} ={\it b} = 19.9598(2) \AA, {\it c} = 16.0609(2)
\AA$\;$ for {Cr$_7$Ni}; {\it a} ={\it b} = 19.9375(2) \AA, {\it c}
= 16.1079(2) \AA$\;$  for {Cr$_7$Mn} and {\it a} ={\it b} =
19.8973(2) \AA, {\it c}  = 16.0227(2) \AA$\;$ for {Cr$_7$Zn}. The
magnetic rings have a diameter of about 10 \AA$\;$ and the metal
ions have the shape of an almost regular planar octagon (Figure
\ref{Cr7Mstructure}).

INS measurements have been performed at the Institute
Laue-Langevin in Grenoble (France), with the direct-geometry
time-of-flight spectrometer IN5, using about $2$ grams of each
compound. A flat disk geometry has been chosen for the aluminum
sample holder ($1.1$ mm thickness, $5$ cm  diameter), in order to
reduce multiple-scattering background. Neutron spectra have been
recorded with the sample kept at different temperatures, between T
= $2$ K and T = $12$ K, inside a standard liquid-$^{4}$He
cryostat. Solid angle and detector efficiency calibrations have
been performed using the spectrum of a vanadium metal sample.
Neutron incident wavelengths $\lambda = 4$ \AA\;, $5$ \AA\; and
$9$ \AA\; were used, corresponding to instrumental resolutions at
the elastic peak of $0.147$ meV, $0.091$ meV, and $0.019$ meV
respectively. The three sets of measurements allowed us to cover
an energy-transfer range from $\hbar \omega \sim 0.04$ meV to
about $4$ meV. The angular interval spanned by the detector banks
corresponds to scattering vector amplitudes varying between $Q
\sim 0.55$ \AA$^{-1}$\; and $2.6$ \AA$^{-1}$\; at $\hbar \omega
\sim 1$ meV, and between $Q \sim 0.85$ \AA$^{-1}$\; and $2.2$
\AA$^{-1}$\; at $\hbar \omega \sim 3.7$ meV (for $\lambda = 4$
\AA\;).

\section{Spin Hamiltonian and INS Magnetic Cross Section}

The investigated systems are ensembles of independent identical
magnetic units, each one described by the Hamiltonian operator:

\begin{eqnarray}
H&=&\sum_{i=1}^{6} J_{Cr-Cr}{\bf s}_{i}\cdot {\bf s}_{i+1} +
J_{Cr-M}({\bf s}_{7}\cdot {\bf s}_{8}+{\bf s}_{8}\cdot {\bf s}_{1})+
\nonumber \\
&+&\sum_{i=1}^{8}{\bf s}_{i}\cdot {\bf D}_{i}\cdot {\bf s}_{i} +
\sum_{i<j=1}^{8}{\bf s}_{i}\cdot {\bf D}_{ij}\cdot {\bf
s}_{j}.\label{eq:H}
\end{eqnarray}

In the above equation we assume that sites $i = 1-7$ of the
octanuclear ring are occupied by $Cr^{3+}$ ions, and site $i=8$ is
filled by the M dication. The first term is the isotropic,
nearest-neighbor Heisenberg-Dirac-Van Vleck exchange interaction,
with $J_{Cr-Cr}$ and $J_{Cr-M}$ being the exchange integrals for
Cr-Cr and Cr-M pairs, respectively. The second term in Eq.
\ref{eq:H} describes the local crystal-field interaction, whereas
the third term gives the dipole-dipole intra-cluster interaction,
evaluated within the point-dipole approximation. Assuming the
${\bf \hat{z}}$-axis along the perpendicular to the ring plane,
the second-order local anisotropy is expected to be dominated by
the axial terms $d_{i}$, with smaller rhombic terms $e_{i}$:

\begin{eqnarray}
\sum_{i=1}^{8}{\bf s}_{i}\cdot {\bf D}_{i}\cdot {\bf s}_{i}&=&
\nonumber \\
\sum_{i=1}^{8}{d_i [s_{z,i}^2 - \frac{1}{3} s_i(s_i+1)]}&+&
\sum_{i=1}^{8}{e_i [s_{x,i}^2 - s_{y,i}^2]}.\label{eq:zfs}
\end{eqnarray}
With $s_{i}\leq 3/2$, no fourth-order local anisotropy must
be considered.

As pointed out by Waldmann \cite{Waldmann03b}, the INS
cross-section described in \cite{Borras99} for polycrystalline
samples of molecular nanomagnets is not of general validity. In
particular, it may fail to describe correctly the scattering from
systems with large magnetic anisotropy and does not properly take
into account intramolecular interference effects. Here we use the
following expression, which
is obtained by averaging the
magnetic dipole INS cross-section \cite{Marshall71} with respect
to the possible orientations of the scattering vector ${\bf Q}$
\cite{Waldmann03b}:

\begin{equation}
\frac{\partial^2 \sigma}{\partial \Omega \partial \omega} =
\frac{A}{N_{m}}
\frac{k_{f}}{k_{0}}e^{-2W}
\sum_{n,m} \frac{e^{-\beta E_n}}{Z} I_{nm}(Q) \delta(\hbar
\omega-E_m+E_n),
\end{equation}

\noindent where $A = 0.29\;$barn, $N_{m}$ is the number of
magnetic ions, $Z$ is the partition function, ${\bf k}_{f}$ and
${\bf k}_{0}$ are the wavevectors of the scattered and incident
neutrons, $\exp(-2W)$ is the Debye-Waller factor, ${\bf Q} = {\bf
k}_{0}-{\bf k}_{f}$ is the scattering vector, $E_{n}$ is the
energy of the generic spin state $|n \rangle$, and the function
I$_{nm}$(Q) is defined as:

\begin{widetext}
\begin{eqnarray}
&&\lefteqn{I_{nm}(Q)=\sum_{i,j} F^*_i(Q) F_j(Q) \times
\left\{\frac{2}{3}
[j_0(QR_{ij})+C_0^2 j_2(QR_{ij})]\tilde{s}_{z_i}\tilde{s}_{z_j}+ \right. \nonumber} \\
&&\left. \frac{2}{3}[j_0(QR_{ij})-
  \frac{1}{2}C_0^2 j_2(QR_{ij})](\tilde{s}_{x_i}\tilde{s}_{x_j}
+\tilde{s}_{y_i}\tilde{s}_{y_j})+ \frac{1}{2}
j_2(QR_{ij})\left[C_2^2(\tilde{s}_{x_i}\tilde{s}_{x_j}-\tilde{s}_{y_i}\tilde{s}_{y_j})+
C^2_{-2}(\tilde{s}_{x_i}\tilde{s}_{y_j}
+\tilde{s}_{y_i}\tilde{s}_{x_j})\right]+ \right.\nonumber \\
&&\left.
j_2(QR_{ij})\left[C_1^2(\tilde{s}_{z_i}\tilde{s}_{x_j}+\tilde{s}_{x_i}\tilde{s}_{z_j})+
C^2_{-1}(\tilde{s}_{z_i}\tilde{s}_{y_j}+\tilde{s}_{y_i}\tilde{s}_{z_j})\right]\right\},
\label{eq:cs}
\end{eqnarray}
\end{widetext}

\noindent where $F(Q)$ is the magnetic form factor, ${\bf R}_{ij}$
gives the relative position of ions $i$ and $j$,

\begin{eqnarray}
C^2_0&=&\frac{1}{2}[3 (\frac{R_{ijz}}{R_{ij}})^2-1] \nonumber \\
C^2_2&=&\frac{R_{ijx}^2-R_{ijy}^2}{R_{ij}^2} \nonumber \\
C^2_{-2}&=&2 \frac{R_{ijx}R_{ijy}}{R_{ij}^2} \nonumber \\
C^2_1&=&\frac{R_{ijx}R_{ijz}}{R_{ij}^2} \nonumber \\
C^2_{-1}&=&\frac{R_{ijy}R_{ijz}}{R_{ij}^2},\label{eq:csa}
\end{eqnarray}

and

\begin{equation}
\tilde{s}_{\alpha_i}\tilde{s}_{\gamma_j}=\langle n \vert
s_{\alpha_i} \vert m\rangle \langle m \vert s_{\gamma_j}\vert n
\rangle  \;\;\; (\alpha,\gamma=x,y,z).
\end{equation}

Eq. \ref{eq:cs} is the explicit form
of the formula given by Waldmann in
\cite{Waldmann03b}.
Besides being valid whatever the symmetry and
amplitude of
the magnetic anisotropy, Eq. \ref{eq:cs} can be easily implemented in
a numeric code.

The parameters of the spin Hamiltonian (Eqs. 1-2) have been
determined from a best fit of the neutron spectra based on
calculations of Eq. \ref{eq:cs}. We recall that from the INS selection
rules only transitions with $\Delta S = 0,\pm 1$ and
$\Delta M = 0,\pm 1$ have non-zero intensity. In the
fitting procedure, a Gaussian lineshape is associated with each
allowed transition, with a Full Width at Half Maximum (FWHM) equal
to the instrumental resolution and an area proportional to the
calculated transition strength. For high-energy transitions, a
broadening of the final state reflecting lifetimes effects has
been assumed. The parameters defining the Hamiltonian have then been
varied until the best agreement between calculated and
experimental spectra was obtained.

The anisotropic part of the spin Hamiltonian can be written in
terms of irreducible tensor operators (ITO) of rank $k=2$, and
therefore mixes states with different $S$ and $M$, or at least
states with different $S$ if the anisotropy is purely axial. Therefore
the
total spin $S$ is not a good quantum number and the
total Hamiltonian cannot be diagonalized within each
$(2S+1)$-dimensional block. This difficulty can be overcome by the
procedure proposed in \cite{Liviotti02a} and used to evaluate the
mixing between the lowest lying $S$ multiplets in the Cr$_{8}$
ring \cite{Carretta03a}.

First, the minimum ensemble of spin manifolds required to
reproduce the INS cross-section at high-energy transfer is
determined assuming isotropic exchange only. Then the complete
Hamiltonian is diagonalized in the corresponding subspace. In this
way one obtains the energy spectrum and the spin states $|n
\rangle$ as linear combinations of basis vectors
$|(\tilde{S})SM\rangle$ labelled by the set of intermediate spin
states $(\tilde{S})$, with coefficients $\langle (\tilde{S})S M|n
\rangle$.

A stringent test for the spectroscopic assignment of the observed
transitions is provided by the $Q$ dependence of their intensity,
which is essentially determined by the geometry of the cluster and
the composition of the spin wavefunctions. This dependence can be
easily measured with properly calibrated detectors at different
scattering angles, and the results can be compared with
calculations using Eq. \ref{eq:cs}.

\section{Experimental Results}

\subsection{Cr$_{7}$Ni}

\begin{figure}[ht]
\includegraphics[width=8cm]{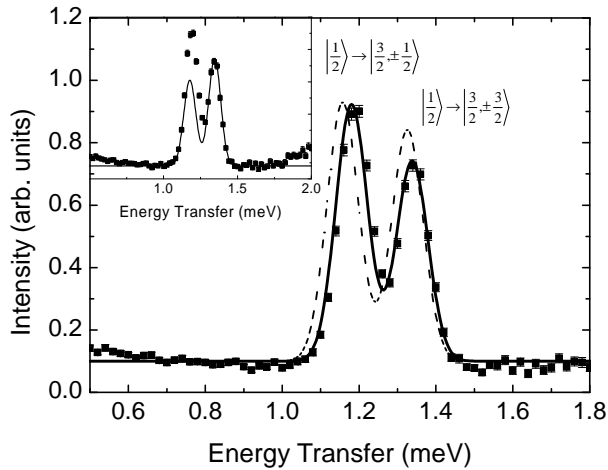}
\caption{Low energy transfer INS response for Cr$_{7}$Ni at $2$ K
($\lambda$ = 5 \AA). The displayed intensity is the sum for the
whole detector bank. Background from the sample holder has been
subtracted. Solid line: intensity calculated for the model
Hamiltonian described in the text, taking into account S-mixing;
broken line: calculated response with S-mixing set to zero. Inset:
intensity calculated with the approximate formula for the
cross-section given in ref.\cite{Borras99} (solid line).}
\label{Cr7Ni5A}
\end{figure}

Figure \ref{Cr7Ni5A} shows the angle-integrated INS intensity
recorded at T = $2$ K for the for Cr$_{7}$Ni sample, with IN5
operated at $\lambda$ = 5 \AA. The two peaks emerging at $1.19$
meV and $1.34$ meV correspond to transitions between the $|S=1/2,
M = \pm 1/2\rangle$ ground state and the anisotropy split
sublevels of the $|S=3/2\rangle$ first excited spin state,
$|S=3/2, M = \pm 1/2\rangle$ and $|S=3/2, M = \pm 3/2\rangle$.
The solid line represents the intensity corresponding to the
Hamiltonian given by Eqs. (1-2), with parameters $J_{Cr-Cr} =
1.46$ meV, $J_{Cr-Ni} = 1.69$ meV, $d_{Cr} = -0.03$ meV, and
$d_{Ni} = -0.6$ meV. The $d_{Cr}$ value is the one determined for
the Cr$_{8}$ ring \cite{Carretta03a} and was kept fixed in the
fitting procedure. As the INS response appears to be quite
insensitive to the in-plane anisotropy, the non-axial part of the
Hamiltonian has been neglected. With the parameters determined in
this work, the ground state composition is dominated by the
$|S=1/2\rangle$ component, with a small $|S=3/2\rangle$
contamination (about $1 \%$); the first excited state is an almost
pure $|S=3/2\rangle$ multiplet with easy-plane zero-field
splitting. The INS cross-section has been calculated using the
expression given by Eq. \ref{eq:cs}, integrated over the $Q$
interval corresponding to the experimental conditions. Although
quantitatively small, S-mixing must be taken into account to
reproduce correctly the intensity ratio of the observed doublet
(the intensity calculated with S-mixing neglected is shown in
Figure \ref{Cr7Ni5A} by the dashed line).

\begin{figure}[ht]
\begin{center}
\includegraphics[width=7.5cm]{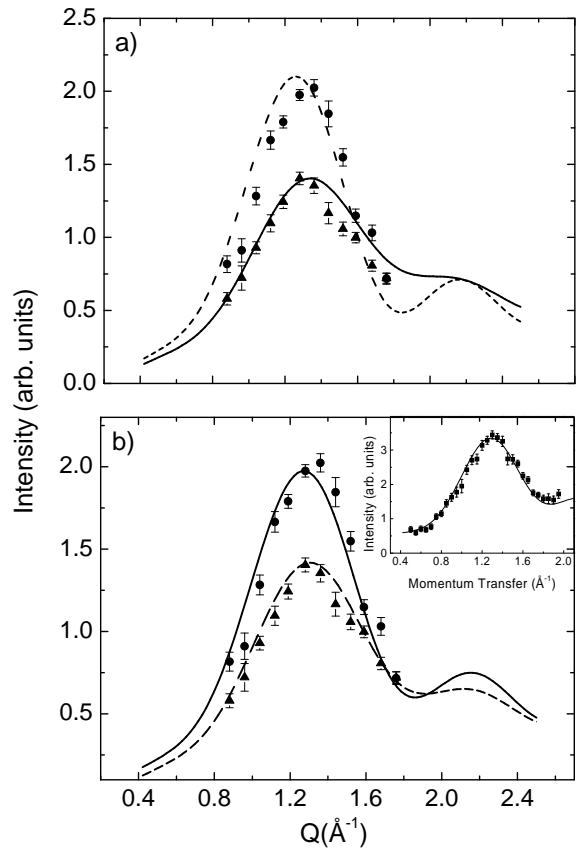}
\caption{Cr$_{7}$Ni INS intensity of the peaks at $1.19$ meV
(circles) and $1.34$ meV (triangles) as a function of the
scattering vector amplitude, $Q$. Data have been obtained with
incident wavelength $\lambda = 5$ \AA\; and sample temperature
$T=2$ K. Calculated curves are represented by solid lines (to be
compared with filled circles) and dashed lines (to be compared
with triangles). Panel a) shows the result obtained with the INS
cross-section as reported in \cite{Borras99}, panel b) shows the
output of Eq. \ref{eq:cs}. In both cases $J_{Cr-Cr} = 1.46$ meV,
$J_{Cr-Ni} = 1.69$ meV, $d_{Cr} = -0.03$ meV, and $d_{Ni} = -0.6$
meV. The inset in panel b) shows the INS response integrated from
$1.1$ and $1.4$ meV as a function of Q, compared with the
calculated curve.} \label{Cr7NiQdep}
\end{center}
\end{figure}

In the inset of Figure \ref{Cr7Ni5A}, the experimental data are
compared with the curve calculated by the approximate INS
cross-section reported in ref. \cite{Borras99}, and the
Hamiltonian parameters quoted above. Due to the large anisotropy
at the Ni site, the Borras-Almenar et al. approximation is not
adequate in the present case, and its use instead of Eq.
\ref{eq:cs} would lead to a completely wrong estimate of the
Hamiltonian parameters. This difference is more evident when
comparing the $Q$ dependence of the transition intensities with
theoretical estimates provided by the two formulae. As shown in
Figure \ref{Cr7NiQdep}a, the intensity of the transition involving
the $|S=3/2, M = \pm 1/2\rangle$ excited level at $1.19$ meV is
underestimated over a large $Q$ interval by the expression given
in \cite{Borras99}, whilst the intensity of the transition towards
the $|S=3/2, M = \pm 3/2\rangle$ state at $1.34$ meV is
overestimated by more than a factor two. In comparison, an
excellent agreement between calculated curves and observed data is
obtained when using Eq. \ref{eq:cs} (Figure \ref{Cr7NiQdep}b).

\begin{figure}[ht]
\includegraphics[width=7.5cm]{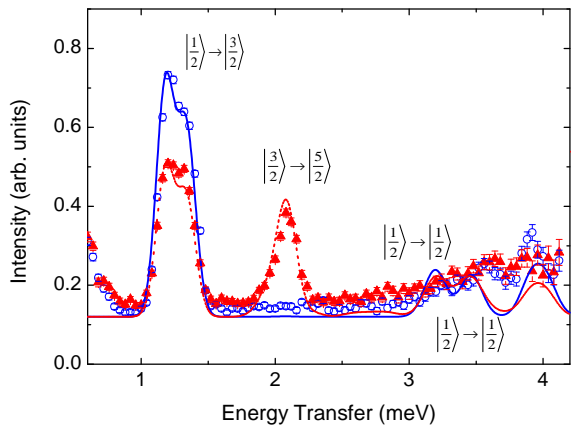}
\caption{Cr$_{7}$Ni INS spectra collected with $\lambda$ = 4 \AA\;
at $T = 2$ K (circles) and $T = 12$ K (triangles). Smooth lines
represent the spectra calculated from eigenvalues and eigenvectors
of the Hamiltonian given in Eqs.\ref{eq:H}-\ref{eq:zfs}, by
associating to each transition a Gaussian lineshape with a height
proportional to the calculated probability and a width of $147$
$\mu$eV, corresponding to the experimental resolution.}
\label{Cr7Ni4A}
\end{figure}

Excitations involving spin manifolds with higher energy can be
observed by reducing the incident wavelength. Results obtained
with $\lambda = 4$ \AA\; at T = $2$ K and T = $12$ K are shown in
Figure \ref{Cr7Ni4A}. At T = 2 K, weak excitations are observed
between  3 and 4 meV. They correspond to transitions from the
ground state to $|S=1/2\rangle$ and $|S=3/2\rangle$ excited
states. When the sample is warmed at $12$ K, a strong peak appears
at $2.08$ meV. This excitation is attributed to the transition
from the first excited $|S=3/2\rangle$ state to a $|S=5/2\rangle$
state lying at $3.31$ meV. The calculated spectra, represented by
the smooth lines in Figure \ref{Cr7Ni4A}, compare very favorably
with the experiment and confirm the validity of the proposed set
of parameters. The ratio $J_{Cr-Ni}/J_{Cr-Cr} = 1.16$ is somewhat
larger than the recent estimate based on the fit of heat capacity
and torque magnetometry data \cite{Troiani04}. In addition the
weighted average axial anisotropy parameter $d = -0.045$ meV
\cite{nota} is about 1.7 times larger than previously reported,
because of the very large anisotropy of the Ni ion that was not
estimated in previous works.

\subsection{Cr$_{7}$Zn}

\begin{figure}[ht]
\includegraphics[width=7.5cm]{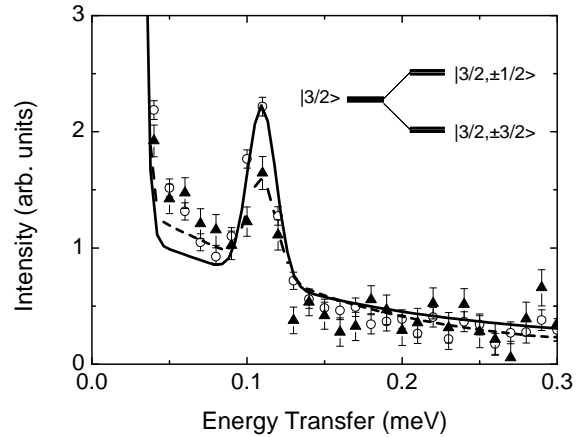}
\caption{Cr$_{7}$Zn high-resolution INS spectra collected with
$\lambda$ = 9 \AA\; at $T = 2$ K (circles) and $T = 10$ K
(triangles). The peak at $0.11$ meV is an intra-multiplet
transition involving the anisotropy split components of the
$|S=3/2>$ ground state. Solid ($2$ K) and dashed ($10$ K) lines
are intensities calculated for a purely axial, easy plane magnetic
anisotropy of the Cr ions ($d_{Cr}$ = $-0.028$ meV). The elastic
peak and a quasielastic contribution have been included in the
best fit procedure.} \label{Cr7Zn9A}
\end{figure}

Substitution of a diamagnetic Zn$^{2+}$ ion for one Cr$^{3+}$ in
the AF Cr$_{8}$ ring results in a $|S=3/2 \rangle$ spin ground
state, whose anisotropy splitting can be observed by
high-resolution INS experiments. Figure \ref{Cr7Zn9A} shows the
spectra recorded on IN5 with incident wavelength $9$ \AA\; (energy
resolution at the elastic peak of $19$ $\mu$eV), with the sample
kept at $2$ and $10$ K.

Assuming crystal-field parameters similar to those determined in
the $Cr_8$ parent compound, we expect an easy-axis anisotropy for
the $S=3/2$ ground manifold and an easy-plane splitting for the
lowest $S=5/2$ multiplet. In this hypothesis, the peak observed at
about $0.11$ meV must be attributed to the $|S=3/2,M=\pm 3/2>
\rightarrow |S=3/2,M =\pm 1/2>$ transition, as only the lowest
manifold is thermally populated at T = $2$ K. The solid and broken
lines give the calculated intensity for $T = 2$ and 10 K assuming
a purely axial local crystal field for the Cr ions, with $d_{Cr}$
= $-0.028$ meV ($d_{Zn} = 0$). As for the Ni case, the INS spectra
are not sensitive to the in-plane component of the anisotropy and
again we consider only the axial part of $H$.

\begin{figure}[ht]
\includegraphics[width=7.5cm]{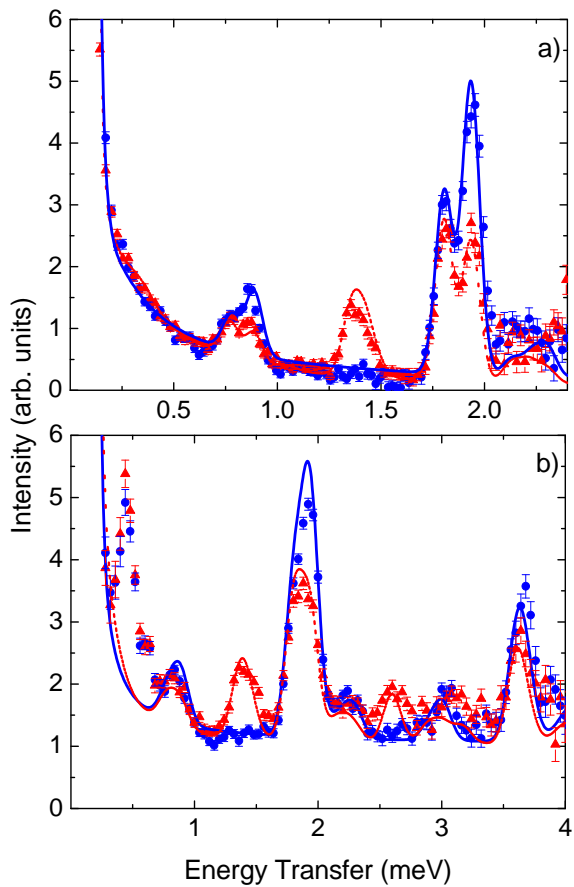}
\caption{INS spectra for Cr$_{7}$Zn at $2$ K (circles) and $12$ K
(triangles), measured with incident wavelength of (a) $5$ \AA\;
and (b) $4$ \AA\;. The $0.4$ meV peak in panel b is spurious.
Smooth lines are best fits of the experimental data to a
superposition of Gaussian line-shapes associated to allowed
transitions. Each Gaussian contribution has an area proportional
to the transition strength calculated with Eq. \ref{eq:cs}
assuming isotropic exchange ($J_{Cr-Cr} = 1.43$ meV, $J_{Cr-Zn} =
0$) and axial local anisotropy ($d_{Cr} = -0.028$ meV).}
\label{Cr7Zn4-5A}
\end{figure}

As shown in Figure \ref{Cr7Zn4-5A}, intermultiplet transitions at
higher energy transfer are observed with incident wavelengths of
$4$ and $5$ \AA. The peaks appearing  in the $T = 2$ K spectrum
are due to transitions from the split $|S=3/2>$ ground state
towards $|S=1/2>$ ($0.83$ meV), $|S=5/2>$ ($1.91$ meV), $|S=1/2>$
and $|S=3/2>$ ($2.25$ meV), $|S=1/2>$ ($3.0$ meV), and unresolved
$|S=5/2>$ and $|S=3/2>$ states ($3.6$ meV).

\begin{figure}
\includegraphics[width=8cm]{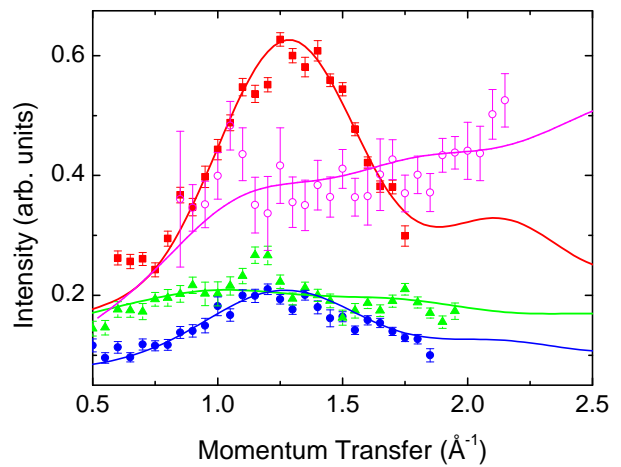}
\caption{$Q$ dependence of the peaks appearing in the Cr$_{7}$Zn
inelastic spectra shown in Figure \ref{Cr7Zn4-5A}, compared with
theoretical estimates (solid lines). Triangles
$|S=3/2>\rightarrow|S=1/2>$ at $0.83$ meV; circles
$|S=1/2>\rightarrow|S=3/2>$ at $1.38$ meV; squares
$|S=3/2>\rightarrow|S=5/2>$ at $1.91$ meV; open circles
$|S=3/2>\rightarrow|S=5/2>$ at $3.6$ meV.} \label{Cr7ZnQdep}
\end{figure}

The spurious peak at $0.4$ meV in panel b is due to neutrons
diffused incoherently by the sample and diffracted by the cryostat
walls. The instrumental resolution in this configurations is not
high enough to resolve the small anisotropy splitting of the
excited levels, so that the energy separation of the doublets
around $0.83$ and $1.91$ meV reflects the splitting of the ground
state. Hot peaks associated to excitations from the first excited
$|S=1/2>$ level appear at $T = 12$ K around $1.39$ and $2.6$ meV.
The solid and broken lines in Figure \ref{Cr7Zn4-5A} are
intensities calculated at the two temperatures assuming an
exchange parameter very similar to the one reported for the
Cr$_{8}$ ring in ref. \cite{Carretta03a}, $J_{Cr-Cr} = 1.43$ meV
for every nearest-neighbor Cr-Cr pairs, $J_{Cr-Zn} = 0$ (as Zn is
diamagnetic) and $d_{Cr} = -0.028$ meV. The agreement with
experimental data is very good and justifies the use of just one
value for the $J_{Cr-Cr}$ exchange integrals, although the use of
a less symmetric magnetic model would improve the fitting at high
energy transfer. The spectroscopic assignments are confirmed by
the $Q$ dependence of the transition intensities, as shown in
Figure \ref{Cr7ZnQdep} where experimental data for some of the
excitations observed are compared with curves calculated with Eq.
\ref{eq:cs}. It is interesting to notice that the general behavior
predicted for cyclic spin clusters by Waldmann \cite{Waldmann03b}
is followed also in the present case where the ring symmetry is
broken by the dication.

\subsection{Cr$_{7}$Mn}

\begin{figure}
\includegraphics[width=8cm]{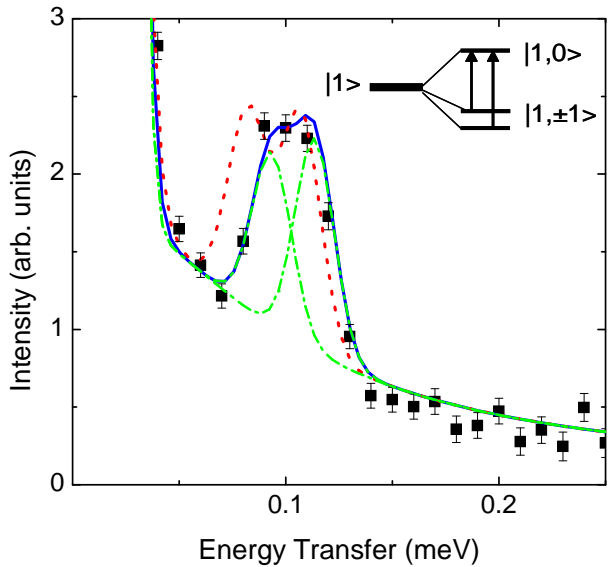}
\caption{Cr$_{7}$Mn high-resolution INS spectra collected with
$\lambda$ = 9 \AA\; at $T = 2$ K. Two intra-multiplet transitions
within the anisotropy split components of the $|S=1>$ ground state
are observed. The solid line is the intensity calculated assuming
both axial and in-plane anisotropy. The dash-dot lines represent
the individual contributions to the broad, unresolved excitation
from the transitions between the split $|S=1, M = \pm 1>$
quasi-doublet and the $|S=1, M = 0>$ singlet. The dot line gives
the intensity calculated with S-mixing set to zero. The elastic
peak and a quasielastic contribution have been included.}
\label{Cr7Mn9A}
\end{figure}

Substitution of one high-spin, $s = 5/2$, Mn$^{2+}$ ion for a
Cr$^{3+}$ in the Cr$_{8}$ ring results in an uncompensated $|S=1>$
ground state. In the presence of purely axial anisotropy, the
zero-field splitting would lead to a quasi-triplet structure, with
an $|S=1, M = \pm 1>$ doublet and an $|S=1, M = 0>$ singlet
separated by an energy proportional to the axial crystal field
parameter. As shown in Figure \ref{Cr7Mn9A}, the $\lambda = 9$
\AA\; high-resolution spectrum recorded at $T = 2$ K shows a broad
excitation that can be fitted by the superposition of two
Gaussians, each with a FWHM equal to the instrumental resolution
($19 \mu$eV) and centered at $0.095$ and $0.11$ meV, respectively.
This is a clear indication of a sizeable rhombic term in the
zero-field splitting Hamiltonian.
By fixing the $d_{Cr}$ parameters to the values determined by INS
for the parent Cr$_{8}$ compound ($d_{Cr} = -0.03$ meV)
\cite{Carretta03a}, the best fit of the data is obtained for
$d_{Mn} = -3$ $\mu$eV (a factor two hundred smaller than the Ni
case) and $e_{Mn} = e_{Cr}=0.14 d_{Mn}$. If the S mixing is
neglected, the ZFS parameters describing the anisotropic splitting
of the $S=1$ ground multiplet can be obtained by projecting the
complete Hamiltonian onto the corresponding subspace. We obtain
$H_{S=1}=D (S_z^2-S(S+1)/3)+E(S_x^2-S_y^2)$ with a rhombicity
$E/D=0.13$. The latter is reduced by S mixing to $E/D=0.1$ (see
Fig. \ref{Cr7Mn9A}).

\begin{figure}[ht]
\includegraphics[width=7.5cm]{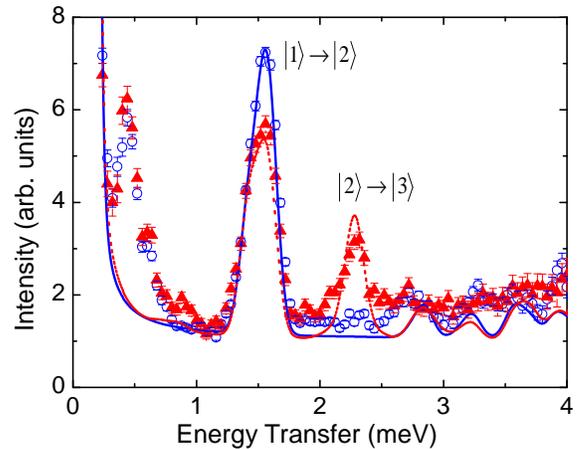}
\caption{Inelastic spectra at $T = 2$ K (circles) and $T = 12$ K
(triangles) obtained for Cr$_{7}$Mn with an incident wavelength
$\lambda = 4$ \AA. The peak at $0.4$ meV is due to spurious
effects. Smooth lines are intensities calculated at $2$ K (solid)
and $12$ K (dashed). The elastic peak and a quasielastic
contribution have been included.} \label{Cr7Mn4A}
\end{figure}

Intermultiplet excitations measured with incident wavelength
$\lambda = 4$ \AA\; are shown in Figure \ref{Cr7Mn4A}. At $T = 2$
K only one strong magnetic peak is observed at $1.6$ meV. A second
magnetic transition appears at $2.3$ meV if the sample is heated
at $12$ K. The two peaks correspond to the transitions involving the ground state
$|S=1>$ and the $|S=2>$ and $|S=3>$ multiplets with minimal energy.

\begin{figure}
\includegraphics[width=7.5cm]{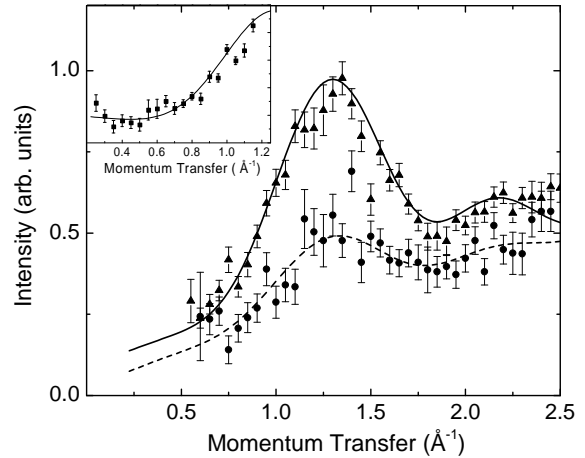}
\caption{$Q$-dependence of the peaks appearing in the Cr$_{7}$Mn inelastic
spectrum at $12$ K, compared with theoretical estimates:
$|S=1>\rightarrow|S=2>$  at $1.6$ meV (triangles and solid line);
$|S=2>\rightarrow|S=3>$ at $2.3$ meV (circles and dashed line). The
$Q$-dependence of the intramultiplet transition at $0.1$ meV ($T = 2$
K) is compared with calculations in the inset.}
\label{Cr7MnQdep}
\end{figure}

Both peaks therefore correspond to excitations within the lowest
rotational-like band and their energy follows the Land\'{e} rule
very closely. In these conditions, only an average exchange
parameter can be estimated. Assuming $J_{Cr-Cr}$ = $J_{Cr-Mn}$ =
$1.43$ meV, calculations provide the spectra shown in Figure
\ref{Cr7Mn4A} by smooth lines. The agreement is satisfactory and a
model with more parameters would not be justified by the data.
Alternatively, by fixing $J_{Cr-Cr}$ = $1.46$ meV as in Cr$_7$Ni,
the same results are obtained with $J_{Cr-Mn}$ = $1.37$ meV. The
assignment of the two transitions to the Land\'{e}-band is
corroborated by the $Q$-dependence of their intensity, shown in
Figure \ref{Cr7MnQdep}. For both peaks a pronounced oscillatory
behavior is observed, whilst an almost flat $Q$ dependence is
expected for transitions to states not belonging to the rotational
band.

\section{Discussion}

\begin{figure}
\bigskip
\bigskip
\includegraphics[width=8.0cm]{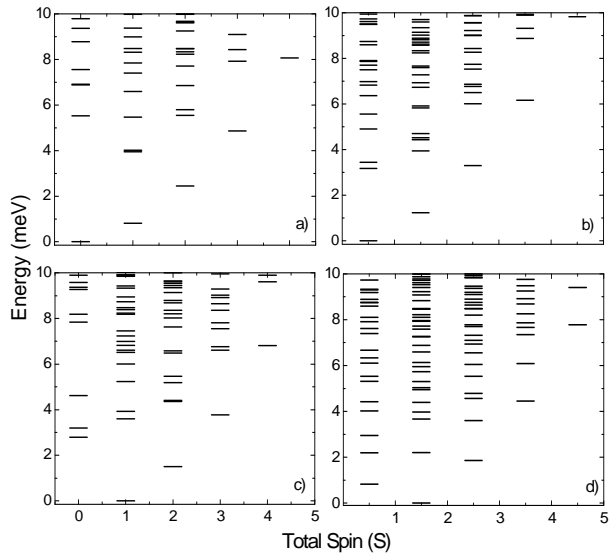}
\caption{Energy of the lowest spin eigenstates as a function of the
total spin $S$ calculated for Cr$_{7}$M magnetic wheels; a) M = Cr, b)
M = Ni, c) M = Mn, d) M = Zn.
Isotropic exchange interactions only are assumed, with parameters
a) $J_{Cr-Cr}$ =
$1.46$ meV \cite{Carretta03a}, b)  $J_{Cr-Cr}$ =$1.46$ meV and $J_{Cr-Ni}$ =
$1.69$ meV,
c) $J_{Cr-Cr}$ =$1.43$ meV and $J_{Cr-Mn}$ = $1.43$ meV, and d)
$J_{Cr-Cr}$ =$1.43$ meV and $J_{Cr-Zn}$ = $0$.} \label{EigCr7M}
\end{figure}

The energy spectra of the heterometallic Cr$_{7}$M magnetic wheels
resulting from our INS investigation are shown in Figure
\ref{EigCr7M} as functions of the total spin, $S$. A parabolic
band, formed by states with minimal energy for each $S$ value, is
easily identified. The levels belonging to this parabolic band
have energies that closely follow the Land\'{e} interval rule,
$E_{S} =
\Delta_{10}[S(S+1)-S_{0}(S_{0}+1)]/[S_{1}(S_{1}+1)-S_{0}(S_{0}+1)]$,
where $S_{0}$ is the spin of the ground state and $\Delta_{10}$ is
the energy of the first excited level, with spin $S_{1}$.
Excitations involving adjacent levels of the parabolic band have
intensities with similar $Q$ dependencies, with an oscillatory
behavior and a pronounced maximum at a $Q$ value related to the
radius of the wheel
\cite{Waldmann01a}. As discussed in
\cite{Waldmann01a,Waldmann03a}, these excitations are related to
the combined quantum rotation of the oppositely oriented total
spin on each
N\'{e}el sublattice of the AF wheel \cite{Anderson52}.

Effects due to the mixing of different spin multiplets have been
considered. Such effects proved to be important to correctly
reproduce the energy and intensity of magnetic excitations in the
neutron spectra. An interesting difference between the spin
wavefunctions of the parent Cr$_{8}$ ring and those of the
heterometallic derivatives concerns their symmetry. In Cr$_{8}$
the ground state and the first excited state belong to different
irreducible representations of the molecule point group, whereas
they have the same symmetry in the substituted wheels. This has
important consequences on the system behavior in the presence of
an external magnetic field. In Cr$_{8}$ the Zeeman splitting of
the $|S=1, \pm 1>$ doublet leads to a crossing with the ground
state at $B = 6.9$ T with no level repulsion and a vanishingly
small Schottky anomaly in the heat capacity \cite{Affronte03}. On
the other hand, according to the results of the diagonalization of
the Hamiltonian Eqs. \ref{eq:H}-\ref{eq:zfs}, anticrossings
between the lowest lying levels are expected in the Cr$_{7}$M
compounds. At the anticrossing conditions, a finite Schottky
anomaly will occour, with an amplitude that depends on the angle
between the applied magnetic field and the easy-axis. Moreover,
the enhancement of $S$-mixing at the level anticrossings will
produce maxima in the torque signal corresponding to oscillations
of the total spin amplitude, as observed for the Mn$[3 \times 3]$
grid \cite{Carretta03b,Waldmann04a}.

In the investigated series, Cr$_{7}$Ni is particularly
interesting. The amount of $S$-mixing in this ring is quite small,
corresponding to about $1 \%$ admixture of $S \neq 1/2$ components
in the ground state. Cr$_{7}$Ni can therefore be considered as an
effective $S = 1/2$ system suitable for the implementation of the
qubit \cite{Troiani04}.
\section{Summary}
Intra-multiplet and inter-multiplet excitations involving spin
manifolds with energy smaller than $4$ meV have been measured with
Inelastic Neutron Scattering in polycrystalline samples of the
heterometallic wheels Cr$_7$M (M = Ni, Mn, Zn). The minimum set of
exchange and local crystal field parameters necessary to describe
the physics of each investigated compound has been determined by
comparing the experimental spectra with theoretical
cross-sections. The results obtained show that chemical
substitution of magnetic ions in a cyclic structure can be used to
tailor the magnetic properties of the wheels, by controlling the
microscopic exchange interaction.
Finally,
the explicit form of the general powder-averaged INS cross-section
given in Ref. \cite{Waldmann03b} and suitable to describe
molecular nanomagnets of any symmetry has been presented.

\begin{acknowledgments}
This work was partly supported by Ministero dell'Universit\`{a} e
della Ricerca Scientifica e Tecnologica, FIRB Project RBNE01YLKN
and by EPSRC(UK). We thank the Institut Laue Langevin, Grenoble,
France for access to the neutron beam facility.

\end{acknowledgments}

\bibliographystyle{apsrev} 
\bibliography{Cr7M}

\begin{thebibliography}{25}
\bibitem[{not()}]{nota}
\bibinfo{note}{This is the value needed to reproduce the observed anisotropy
  splitting of the $|S=3/2\rangle$ manifold by assuming $d_{Ni}=d_{Cr}$.}
\end{thebibliography}
\end{document}